\documentclass[aps,prl,reprint,tightenlines,superscriptaddress,floatfix]{revtex4-1}

\usepackage{graphicx}
\usepackage{amsmath}
\usepackage{amssymb}
\usepackage{color}
\usepackage{dcolumn}
\usepackage{xspace}
\usepackage{attrib}
\usepackage{mathtools}

\usepackage{xspace}
\makeatletter
\DeclareRobustCommand\onedot{\futurelet\@let@token\@onedot}
\def\@onedot{\ifx\@let@token.\else.\null\fi\xspace}
 
\def\ie{{i.e}\onedot}

\makeatother

\newcommand{\nuc}[2]{\hbox{$^{#1}$#2}}

\begin{document}

\title{Neutron single-particle strength in silicon isotopes: constraining the
  driving forces of shell evolution}

\author{S.R.~Stroberg} 
\altaffiliation{Present address: TRIUMF, Vancouver, British Columbia V6T 2A3, Canada}
   \affiliation{National Superconducting Cyclotron Laboratory,
      Michigan State University, East Lansing, Michigan 48824, USA}
   \affiliation{Department of Physics and Astronomy,
      Michigan State University, East Lansing, Michigan 48824, USA}
\author{A.~Gade} 
\affiliation{National Superconducting Cyclotron Laboratory,
      Michigan State University, East Lansing, Michigan 48824, USA}
   \affiliation{Department of Physics and Astronomy,
      Michigan State University, East Lansing, Michigan 48824, USA}
\author{J.A.~Tostevin} 
   \affiliation{Faculty of Engineering and Physical Sciences,
      University of Surrey, Guildford, Surrey, GU2 7XH, United Kingdom}
\author{V.M.~Bader} 
\affiliation{National Superconducting Cyclotron Laboratory,
      Michigan State University, East Lansing, Michigan 48824, USA}
   \affiliation{Department of Physics and Astronomy,
      Michigan State University, East Lansing, Michigan 48824, USA}
\author{T.~Baugher} 
\affiliation{National Superconducting Cyclotron Laboratory,
      Michigan State University, East Lansing, Michigan 48824, USA}
   \affiliation{Department of Physics and Astronomy,
      Michigan State University, East Lansing, Michigan 48824, USA}
\author{D.~Bazin} 
   \affiliation{National Superconducting Cyclotron Laboratory,
      Michigan State University, East Lansing, Michigan 48824, USA}
   \affiliation{Department of Physics and Astronomy,
      Michigan State University, East Lansing, Michigan 48824, USA}
\author{J.S.~Berryman} 
   \affiliation{National Superconducting Cyclotron Laboratory,
      Michigan State University, East Lansing, Michigan 48824, USA}
\author{B.A.~Brown} 
\affiliation{National Superconducting Cyclotron Laboratory,
      Michigan State University, East Lansing, Michigan 48824, USA}
   \affiliation{Department of Physics and Astronomy,
      Michigan State University, East Lansing, Michigan 48824, USA}
\author{C.M.~Campbell}
   \affiliation{Lawrence Berkeley National Laboratory,
       Berkeley, California 94720, USA}
\author{K.W.~Kemper} 
   \affiliation{Department of Physics,
       Florida State University, Tallahassee, Florida 32306, USA}
\author{C.~Langer} 
   \affiliation{National Superconducting Cyclotron Laboratory,
      Michigan State University, East Lansing, Michigan 48824, USA}
   \affiliation{Joint Institute for Nuclear Astrophysics,
      Michigan State University, East Lansing, Michigan 48824, USA}
\author{E.~Lunderberg} 
   \affiliation{National Superconducting Cyclotron Laboratory,
      Michigan State University, East Lansing, Michigan 48824, USA}
   \affiliation{Department of Physics and Astronomy,
      Michigan State University, East Lansing, Michigan 48824, USA}
\author{A.~Lemasson} 
   \affiliation{National Superconducting Cyclotron Laboratory,
      Michigan State University, East Lansing, Michigan 48824, USA}
\author{S.~Noji} 
   \affiliation{National Superconducting Cyclotron Laboratory,
      Michigan State University, East Lansing, Michigan 48824, USA}
\author{T. Otsuka} 
   \affiliation{Department of Physics, University of Tokyo,
       Hongo, Bunkyo-ku, Tokyo 113-0033, Japan}
   \affiliation{Center for Nuclear Physics, University of Tokyo,
       Hongo, Bunkyo-ku, Tokyo 113-0033, Japan}
   \affiliation{National Superconducting Cyclotron Laboratory,
      Michigan State University, East Lansing, Michigan 48824, USA}
\author{F.~Recchia} 
   \affiliation{National Superconducting Cyclotron Laboratory,
      Michigan State University, East Lansing, Michigan 48824, USA}
\author{C.~Walz} 
   \affiliation{National Superconducting Cyclotron Laboratory,
      Michigan State University, East Lansing, Michigan 48824, USA}
\author{D.~Weisshaar} 
   \affiliation{National Superconducting Cyclotron Laboratory,
      Michigan State University, East Lansing, Michigan 48824, USA}
\author{S.~Williams}
   \affiliation{National Superconducting Cyclotron Laboratory,
      Michigan State University, East Lansing, Michigan 48824, USA}

\begin{abstract}
Shell evolution is studied in the neutron-rich silicon isotopes \nuc{36,38,40}{Si}
using neutron single-particle strengths deduced from one-neutron knockout reactions.
Configurations involving neutron excitations across the $N=20$ and $N=28$ shell
gaps are quantified experimentally in these rare isotopes. Comparisons with
shell model calculations show that the tensor force, understood to drive the
collective behavior in \nuc{42}{Si} with $N=28$, is already important in
determining the structure of \nuc{40}{Si} with $N=26$.
New data relating to cross-shell excitations provide the first quantitative
support for repulsive contributions to the cross-shell $T=1$ interaction
arising from three-nucleon forces.

\end{abstract}
\maketitle

The atomic nucleus is a fermionic many-body quantum system composed of
strongly-interacting protons and neutrons. Large stabilizing energy gaps,
separating clusters of single-particle states, provide the cornerstone
for the nuclear shell model, one of the most powerful tools available for
describing the structure of atomic nuclei. In the simplest version of
the shell model, empirical shell gaps at the {\em magic} nucleon numbers
2, 8, 20, 28, 50, 82, and 126 are reproduced when assuming that the nucleons
experience, predominantly, a mean-field potential with an attractive
one-body spin-orbit term.

In rare isotopes, with imbalanced proton and neutron numbers, significant
modifications have been observed. Here, new shell gaps develop and the
conventional gaps at the magic numbers can collapse. Understanding this
observed evolution is key to a comprehensive description of atomic nuclei
across the nuclear chart. Detailed studies of the evolution of shell
structure with proton number ($Z$) or neutron number ($N$), e.g. \cite{Sorlin2008},
probe the effects of particular components of the complex interactions
between nucleons: such as the spin-isospin \cite{Otsuka2001} and tensor
\cite{Otsuka2005} two-body terms, and three-body force terms \cite{Zuker2003,Otsuka2010a}.
The need to include such terms in the nuclear interaction has been demonstrated
by their robust effects that become amplified at large isospin
\cite{Otsuka2005,Otsuka2010a} and, without which, features such as driplines
and shell structure may not be reproduced. Clearly, a full treatment of
the nuclear force from its underlying QCD degrees of freedom is very challenging,
and experimental data is essential in helping to identify the most important
degrees of freedom responsible for driving the evolution of nuclear properties.

Here, we present data for the silicon ($Z=14$) isotopic chain, a region of
the nuclear chart were rapid shell evolution is at play. \nuc{34}{Si} is known
to exhibit closed-shell behavior while \nuc{42}{Si} shows no indication of an
$N=28$ shell gap \cite{Bastin2007}. Hitherto, observations on the neutron-rich
silicon isotopes, dominated by measurements of collective observables, have been
reproduced by large-scale shell model calculations using phenomenological
effective interactions \cite{Bastin2007,Nowacki2009,Utsuno2012,Caurier2014}.
To assess the theoretical description of the evolving shell structure it is
also critical to investigate these nuclei using single-particle observables,
such as the energies and single-particle (spectroscopic) strengths of states
involving the active orbitals at shell gaps.

This Rapid Communication reports a first experimental investigation of observables that reflect
single-neutron degrees of freedom in the \nuc{36,38,40}{Si} isotopes.
Extraction of the presented cross-sections from the data, collected in the measurement reported in
 Ref. \cite{Stroberg2014}, required the development of novel analysis strategies.
The results go beyond those of Ref. \cite{Stroberg2014}, and are interpreted here within a common theoretical framework.
Exclusive one-neutron knockout cross sections, measured
using $\gamma$-ray tagged neutron removal reactions from \nuc{36,38,40}{Si}
projectiles, are used to identify and quantify configurations that involve neutron
excitations across the $N=20$ and $N=28$ shell gaps. Specifically, the partial
cross sections to the lowest-lying $7/2^-$ and $3/2^-$ states, involving the
diminishing $N=28$ gap, and $1/2^+$ and $3/2^+$ states, involving the $N=20$
shell gap, are measured and compared to calculations using shell-model spectroscopic
strengths and eikonal reaction theory. The results (i) track the evolution of the
neutron $f_{7/2}$ and $p_{3/2}$ orbitals at the $N=28$ shell gap, and (ii) quantify
the little-explored neutron excitations, from the $d_{3/2}$ and $s_{1/2}$ $sd$-shell
orbitals, across the $N=20$ gap.

The experiment was performed at the Coupled Cyclotron Facility of the National
Superconducting Cyclotron Laboratory at Michigan State University. Secondary beams
of \nuc{36,38,40}{Si}, produced by fast fragmentation of a \nuc{48}{Ca} primary beam,
impinged on a beryllium target with energies of 100, 95 and 85 MeV/u, respectively.
The one-neutron knockout residues were detected and identified on an event-by-event basis.
Prompt $\gamma$-rays, emitted in-flight from de-excitation of the knockout
residues, were detected with the GRETINA array \cite{Paschalis2013} surrounding the
target position, and were Doppler-corrected event-by-event.

\begin{figure}
\includegraphics[width=1.0\columnwidth]{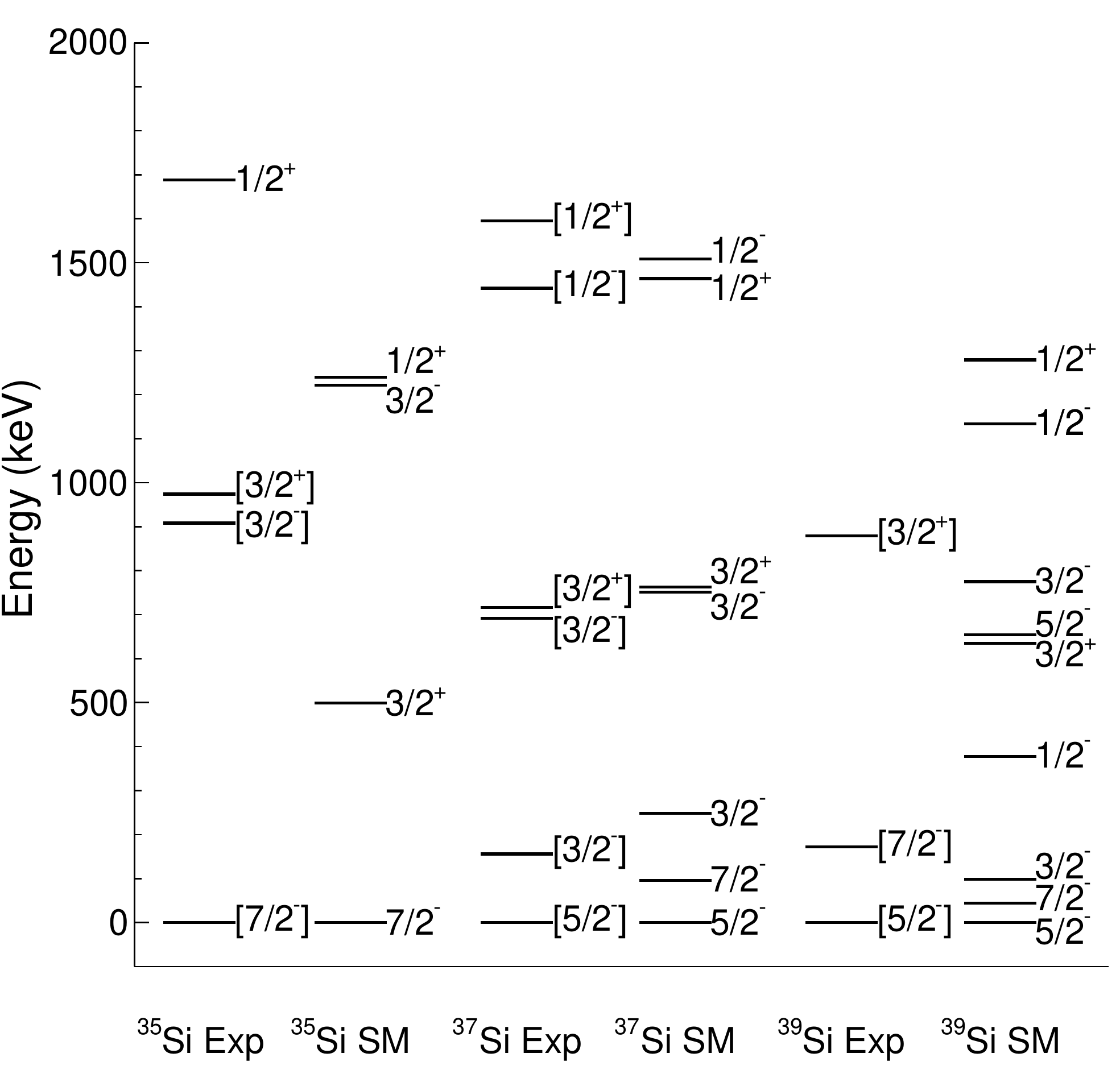}
\caption{Experimental level schemes for low-lying states in \nuc{35}{Si},
\nuc{37}{Si} and \nuc{39}{Si}, compared with shell model calculations using
the SDPF-MU effective interaction \cite{Stroberg2014}.}
\label{fig:levelschemes}
\end{figure}

The level schemes of the knockout residues were constructed based on $\gamma\gamma$
coincidences, energy sums, and intensity balances. These are summarized in
Fig.\ \ref{fig:levelschemes}. Spin-parity assignments were made with the aid of the
parallel momentum distributions of the residues in comparison with theoretical
distributions calculated in an eikonal model according to the formalism of Ref.
\cite{Bertulani2006}. Full details of the experiment, data analysis
and the spin-parity assignments can be found in Ref. \cite{Stroberg2014}.

\begin{figure}
\includegraphics[width=1.0\columnwidth]{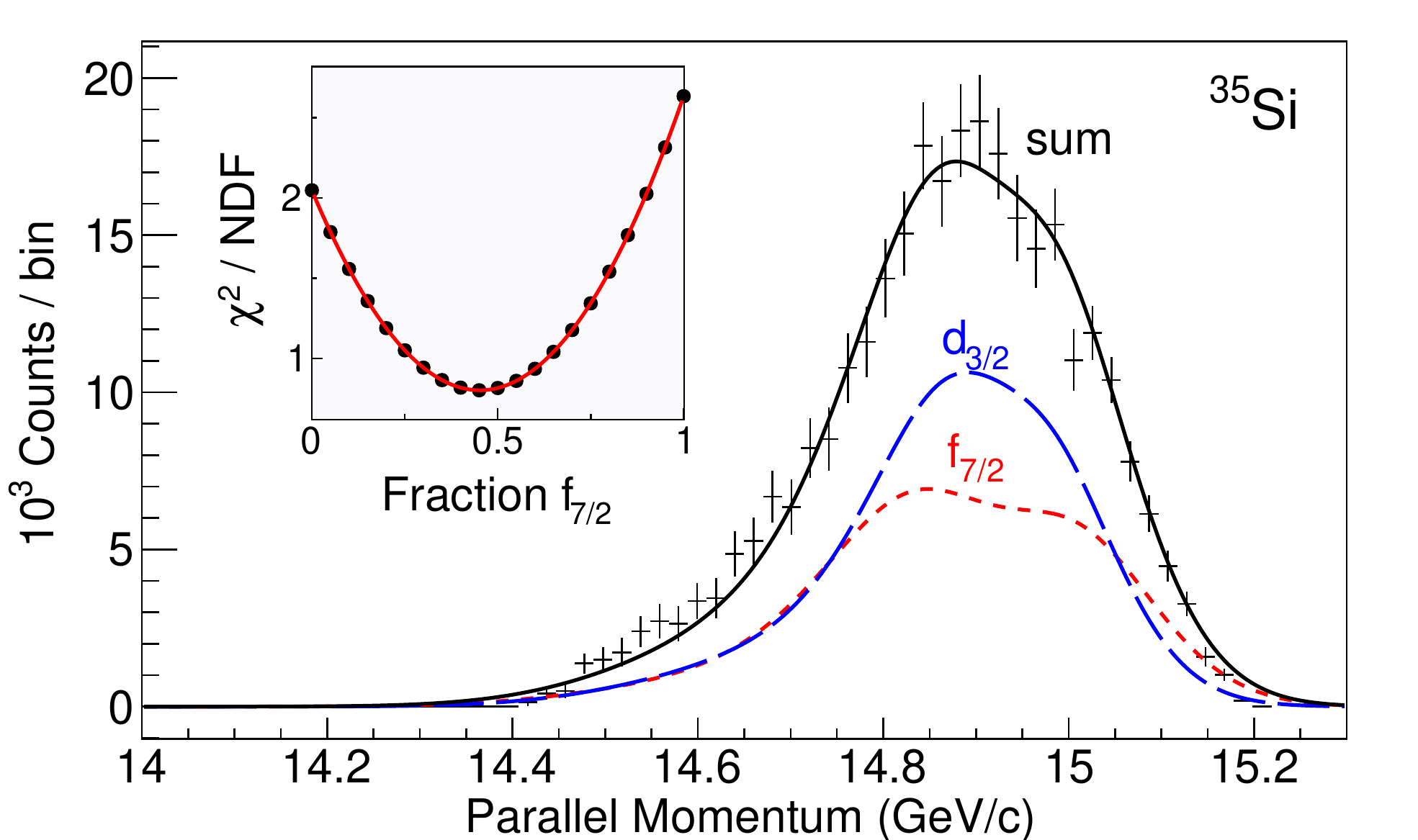}
\caption{(color online) Momentum distribution of the \nuc{35}{Si} residues
produced in the ground state and any isomers following one neutron knockout
from \nuc{36}{Si}. The curves show a fit using a linear combination of the
calculated distributions for removal from the $f_{7/2}$ and $d_{3/2}$ orbits.}
\label{fig:Si35gsFit}
\end{figure}

The knockout cross sections to negative parity states, i.e. removal from the
neutron $f_{7/2}$ and $p_{3/2}$ orbitals in \nuc{36,38,40}{Si}, map their
spectroscopic strengths. The experimental and calculated cross sections are
listed in Table \ref{tab:CSfp}. Details of the shell model calculations used
can be found in Ref. \cite{Stroberg2014}. Determining the partial cross sections is
challenging in some cases. For example, population of the \nuc{35}{Si}($7/2^-_1$)
ground state is hindered by the presence of a $3/2^{+}$ isomer, expected to
be strongly populated but which cannot be tagged with prompt in-beam $\gamma$
spectroscopy. We use instead the \nuc{35}{Si} residue momentum distribution
to extract the population fraction. Fig.\ \ref{fig:Si35gsFit} shows the
\nuc{35}{Si} parallel momentum distribution after the subtraction of all events
that decay by prompt $\gamma$ emission. Overlaid is a linear combination
of the theoretical distributions for neutron removal from the $f_{7/2}$ and
$d_{3/2}$ orbits together with the resulting $\chi^2$
fit minimization, giving a $f_{7/2}$ fraction of 0.45(10). The ground state
cross section is estimated from this fraction of the knockout reaction events
with no prompt $\gamma$ decay.

\begin{table}
\caption{Experimental ($\sigma_{exp}$) and calculated ($\sigma_{th}$) one-neutron
knockout cross sections to the lowest $7/2^-$ and $3/2^-$ states in the mass $A_{res}$
residues. The $\sigma_{th}$ use the shell-model spectroscopic factors $C^2S$ and their
center-of-mass correction, $(A_{proj}/A_{res})^3$, and the calculated eikonal model
single-particle cross sections $\sigma_{sp}$. All cross sections are in millibarns.}
\label{tab:CSfp}
\begin{ruledtabular}
\begin{tabular}{ccc D{.}{.}{-1} D{.}{.}{-1}D{.}{.}{-1}D{.}{.}{-1} D{.}{.}{-1}D{(}{(}{-1}}
&&&& \multicolumn{2}{c}{SDPF-MU} & \multicolumn{2}{c}{SDPF-U} \\
$J^\pi_f$   &  $A_{proj}$   & $A_{res}$ &
\multicolumn{1}{c}{$\sigma_{sp}$} &
\multicolumn{1}{r}{$C^2S$} &
\multicolumn{1}{c}{$\sigma_{th}$} &
\multicolumn{1}{r}{$C^2S$} &
\multicolumn{1}{c}{$\sigma_{th}$} &
\multicolumn{1}{c}{$\sigma_{exp}$} \\
\hline
$7/2^-_1$   & 36 & 35  & 15.7                & 1.71 & 29.2  & 1.73 & 29.5  & 23(6) \\
            & 38 & 37  & 15.2                & 2.81 & 46.3  & 2.85 & 46.6  & 47(9) \\
            & 40 & 39  & 15.0                & 3.19 & 51.8  & 3.33 & 53.9  & 49(7) \\
            & 42 & 41  & 15.9\footnotemark[1]& 2.73 & 46.6  & 3.70 & 64.1  &        \\
\hline
$3/2^-_1$   & 36 & 35  & 17.8                & 0.13 &  2.6  & 0.09 &  1.8  &  8(3)  \\
            & 38 & 37  & 20.1                & 0.11 &  2.4  & 0.02 &  0.4  &  9(7) \\
            & 40 & 39  & 21.7\footnotemark[1]& 0.90 & 21.3  & 0.51 & 12.2  & 29(20) \\
            & 42 & 41  & 26.7\footnotemark[1]& 1.72 & 49.4  & 0.03 &  1.0  &        \\
\hline
$3/2^-_2$   & 38 & 37  & 18.8                & 0.27 &  5.5  & 0.27 &  5.6  &  7(3) \\
            & 40 & 39  & 19.8\footnotemark[1]& 0.08 &  1.8  & 0.17 &  3.7  &       \\
            & 42 & 41  & 22.6\footnotemark[1]& 0.19 &  4.6  & 1.02 & 30.5  &       \\
\end{tabular}
\end{ruledtabular}
\footnotetext[1]{Experimental excitation energy not known. $\sigma_{sp}$ is
calculated with the SDPF-MU shell model energy.}
\end{table}

\begin{figure}[ht!]
\includegraphics[width=1.0\columnwidth]{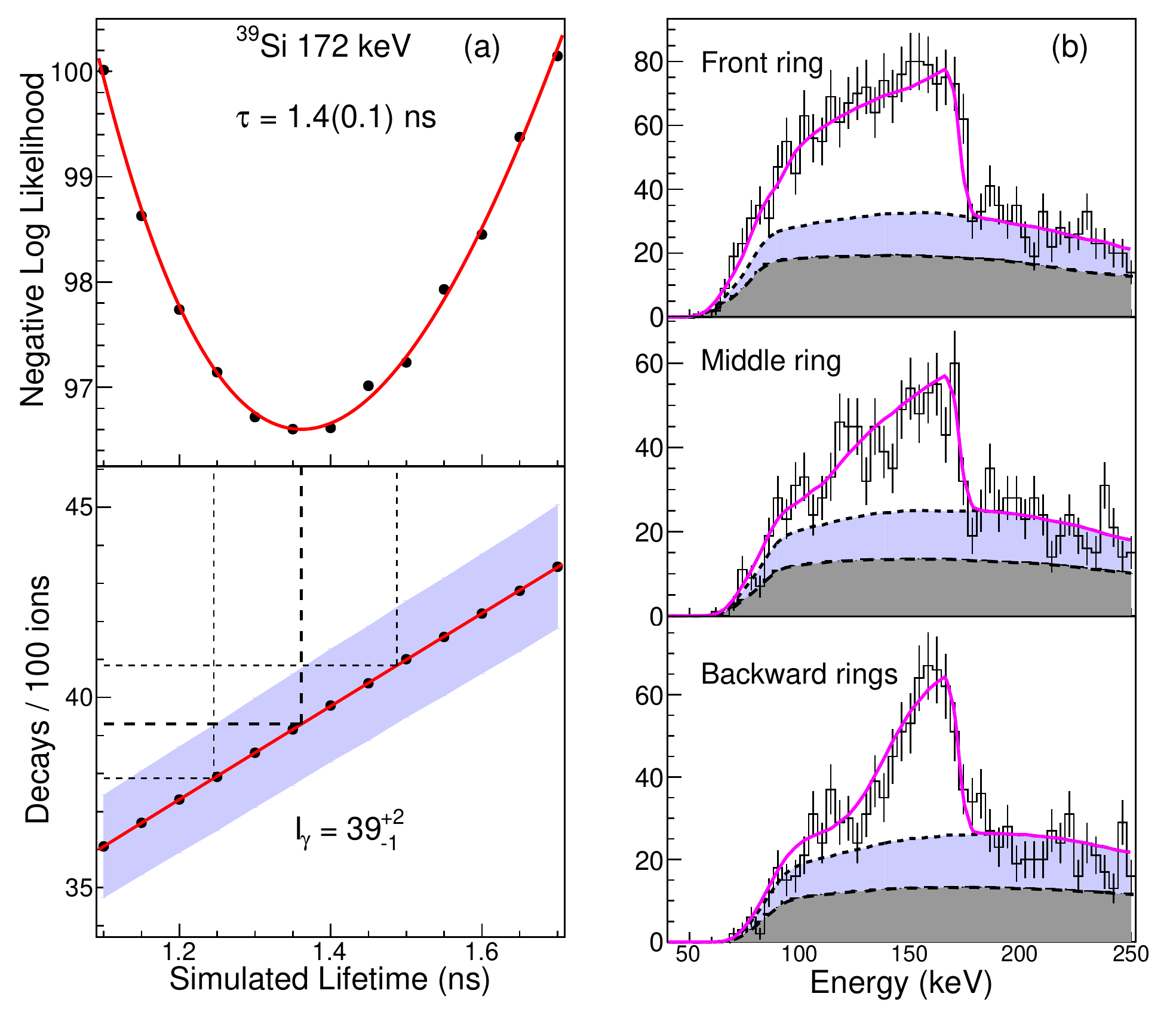}
\caption{(color online) (a) Maximum likelihood fit to the lifetime of the
state in \nuc{39}{Si} decaying by a 172 keV $\gamma$ ray. Only statistical
uncertainties are shown. The lower panel shows the influence of the uncertainty
in lifetime upon the uncertainty in the number of decays, where the error band
shows the uncertainty in the fit. (b) The resulting spectra in the front, middle
and backward rings of GRETINA using the best-fit lifetime. The shaded gray area
indicates the continuous background from target breakup, the blue area indicates
the Compton continuum of all higher-lying peaks, and the magenta curve indicates
the total fit \cite{Stroberg2014}.} \label{fig:Si39LT}
\end{figure}

The \nuc{37}{Si}($3/2^-_1$) and \nuc{39}{Si}($7/2^-_1$) states are nanosecond
isomers. As a result, their depopulating transitions have broad, asymmetric
peak shapes due to the larger uncertainty in the position and velocity of
the decaying fragment and corresponding degradation of the Doppler reconstruction.
This lifetime effect is incorporated into the GEANT4 \cite{Agostinelli2003} simulation
 and a best-fit lifetime
is obtained with a maximum likelihood method. An example is shown in Fig.\ \ref{fig:Si39LT}.
The proximity of these peaks to the $\gamma$-ray detection
threshold results in a dependence of the extracted peak intensity on the
assumed lifetime. This dependence is shown in the lower panel of Fig.\ \ref{fig:Si39LT}(a),
and makes the major contribution to the uncertainty in
these peak intensities. The effect of the lifetime on the peak shape depends
on the polar angle of the emitted $\gamma$ ray. We can confirm that the
simulation reproduces this dependence by dividing the array into three
rings centered near 50, 65 and 90 degrees (labeled front, middle, and
backward in Fig.\ \ref{fig:Si39LT}), and comparing the fit in each ring.
This comparison, in Fig.~\ref{fig:Si39LT}(b), shows satisfactory agreement.

\begin{figure}
\includegraphics[width=1.0\columnwidth]{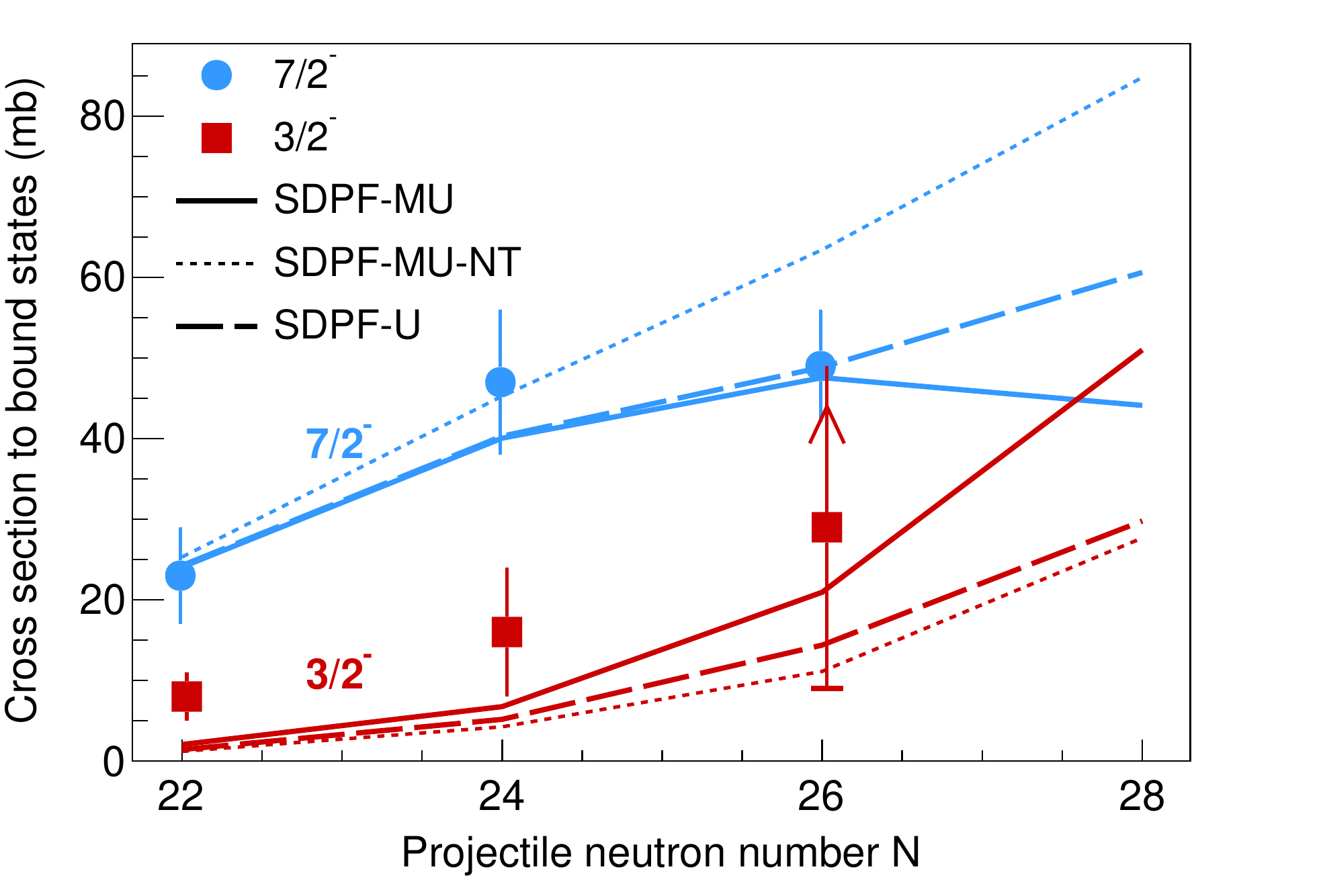}
\caption{(color online) One-neutron knockout reaction partial cross sections to
bound final states with $J^{\pi}=7/2^-$ and $3/2^-$, as a function of the neutron
number of the silicon projectile. The arrow, for $N=26$, indicates a lower limit.
Theoretical predictions using the SDPF-MU (solid lines) and SDPF-U (dashed lines)
effective interactions are also shown. The dotted line (SDPF-MU-NT) results when
the tensor component of the cross-shell $sd$-$fp$ interaction of SDPF-MU is
set to zero. All theoretical curves have been scaled by the observed knockout
reaction systematics (see text).}
\label{fig:CSfp}
\end{figure}

The large uncertainty for the \nuc{39}{Si}($3/2^-_1$) state in Table \ref{tab:CSfp}
is due to several observed transitions which were not placed in the level scheme,
introducing ambiguity in the subtraction procedure described above. The quoted
uncertainty includes the range of possible level schemes which are consistent
with the data. Further, since the second $3/2^-$ state was not identified, the
value shown provides only a lower limit on the bound $p_{3/2}$ strength.
The stated \nuc{37,39}{Si}($7/2^-$) cross sections assume that population of
the predicted $5/2^-_1$ states is small compared to other sources of uncertainty
(the shell model strengths predict cross sections of order 1 mb). The measured
and theoretical cross sections (for the SDPF-MU \cite{Utsuno2012} and SDPF-U
\cite{Nowacki2009} shell model effective interactions) are shown in
Fig.\ \ref{fig:CSfp}. In the region of \nuc{42}{Si}, the tensor component of 
the interaction has been proposed as an important driving force for shell 
evolution \cite{Otsuka2005,Bastin2007}, and so we investigate this with a third set of calculations---denoted
SDPF-MU-NT---obtained by removing the tensor part of the cross-shell $sd$-$fp$
interaction of SDPF-MU. All theoretical cross sections are scaled by an
empirical quenching factor $R(\Delta S)$ obtained from a fit to the knockout
reaction systematics \cite{Gade2008a,Stroberg2014}.

The agreement between the measured $7/2^-$ state cross sections (shown in blue)
and both the SDPF-MU and SDPF-U calculations is excellent. We see that the
effect of the tensor force, as discussed in \cite{Otsuka2005}, becomes important
\footnote{At \nuc{40}{Si}, the effect of the tensor force becomes important 
relative to the theoretical precision, which is approximately 10-20\% from systematics}
already around \nuc{40}{Si}.
In contrast, the $3/2^-$ state cross sections (shown
in red) are markedly under-predicted. This finding is consistent with previous
measurements using one-neutron knockout, from \nuc{30,32}{Mg} \cite{Terry2008}
and \nuc{33}{Mg} \cite{Kanungo2010}, as well as a $(t,p)$ transfer measurement
populating states in \nuc{32}{Mg} \cite{Wimmer2010}. In each of these cases,
an excess of $p_{3/2}$ strength was seen, relative to shell model predictions,
while the $f_{7/2}$ strength was generally consistent with the shell model.
The fact that this discrepancy is observed for different reaction mechanisms,
and only for a particular orbit, suggests that this is a structure effect and
not related to any systematic defect of the reaction theory. As can be seen
from the dotted line in Fig.~\ref{fig:CSfp}, the tensor force does not appear
to have much effect in \nuc{36}{Si} and \nuc{38}{Si}, and so the $p_{3/2}$
discrepancy likely has origins elsewhere.

To clarify the $N=20$ shell closure we also consider the removal of neutrons
from the $d_{3/2}$ and $s_{1/2}$ $sd$-shell orbitals, populating positive-parity
final states. The cross sections for population of bound $3/2^+$ and $1/2^+$
states are listed in Table \ref{tab:CSsd}. The large uncertainty for the
\nuc{35}{Si}($3/2^+_1$) state yield is due to the same isomer effect as was
discussed for the $7/2^-_1$ state. The measured and theoretical (SDPF-MU and
SDPF-U) cross sections are compared in Table~\ref{tab:CSsd}. Both model
calculations, which include $1p-1h$ excitations from the $sd$-shell in the
wave functions of the residual nuclei, over-predict the strength of transitions
from these orbits.

\begin{table}
\caption{Experimental ($\sigma_{exp}$) and calculated ($\sigma_{th}$)
one-neutron knockout cross sections to the lowest $3/2^+$ and $1/2^+$ states
in the mass $A_{res}$ residues. The $\sigma_{th}$ use the shell-model
spectroscopic factors $C^2S$ and their center-of-mass correction,
$(A_{proj}/A_{res})^2$, and the calculated eikonal model single-particle
cross sections $\sigma_{sp}$. All cross sections are in millibarns.}
\label{tab:CSsd}
\begin{ruledtabular}
\begin{tabular}{ccc D{.}{.}{-1} D{.}{.}{-1}D{.}{.}{-1}D{.}{.}{-1} D{.}{.}{-1}D{(}{(}{-1}}
&&&& \multicolumn{2}{c}{SDPF-MU} & \multicolumn{2}{c}{SDPF-U} \\
$J^\pi_f$   &  $A_{proj}$   & $A_{res}$ &
\multicolumn{1}{c}{$\sigma_{sp}$} &
\multicolumn{1}{r}{$C^2S$} &
\multicolumn{1}{c}{$\sigma_{th}$} &
\multicolumn{1}{r}{$C^2S$} &
\multicolumn{1}{c}{$\sigma_{th}$} &
\multicolumn{1}{c}{$\sigma_{exp}$} \\
\hline
$3/2^+_1$   & 36 & 35   & 14.2                 & 3.07 & 46.0  & 2.61 & 39.2  & 29(6) \\
            & 38 & 37   & 13.8                 & 2.79 & 40.6  & 2.19 & 28.4  & 19(2) \\
            & 40 & 39   & 13.3                 & 2.31 & 32.4  & 1.69 & 19.9  & 14(2) \\
\hline
$1/2^+_1$   & 36 & 35   & 21.1                 & 0.96 & 21.3  & 1.00 & 22.3  & 13(1)  \\
            & 38 & 37   & 20.7                 & 0.80 & 17.5  & 0.97 & 19.5  & 10(1) \\
            & 40 & 39   & 22.7\footnotemark[1] & 0.53 & 12.6  & 0.72 & 14.1  &       \\
\end{tabular}
\end{ruledtabular}
\footnotetext[1]{Experimental excitation energy not known. $\sigma_{sp}$
is calculated with the SDPF-MU shell model energy.}
\end{table}

It is very likely that the theoretical over-prediction of $sd$-shell strength and
the aforementioned under-prediction of $fp$-shell strength are related, reflecting
unaccounted-for excitations across the $N=20$ shell gap in the ground states of
the projectiles. Indeed, the present calculations for the projectile ground states
were performed in a $0\hbar \omega$ model space in which the neutron $sd$-shell 
orbits are fixed and fully occupied. So, it is evident that the assumed occupation
of $sd$-shell orbits is too high. In the Monte Carlo shell model calculations of
Ref. \cite{Utsuno1999}, that allowed an arbitrary number of neutron particle-hole
excitations from the $sd$-shell into the lower $fp$ shell, the results, in
\nuc{36,38}{Si}, were an average excess of approximately 0.3$-$0.4 neutrons
compared to normal filling. This reduced $sd$ strength (and additional $fp$
strength) would bring the shell-model predictions into better agreement with
the present data, with the exception of the large $3/2^+$ strength of SDPF-MU.

\begin{figure}[ht]
\includegraphics[width=1.0\columnwidth]{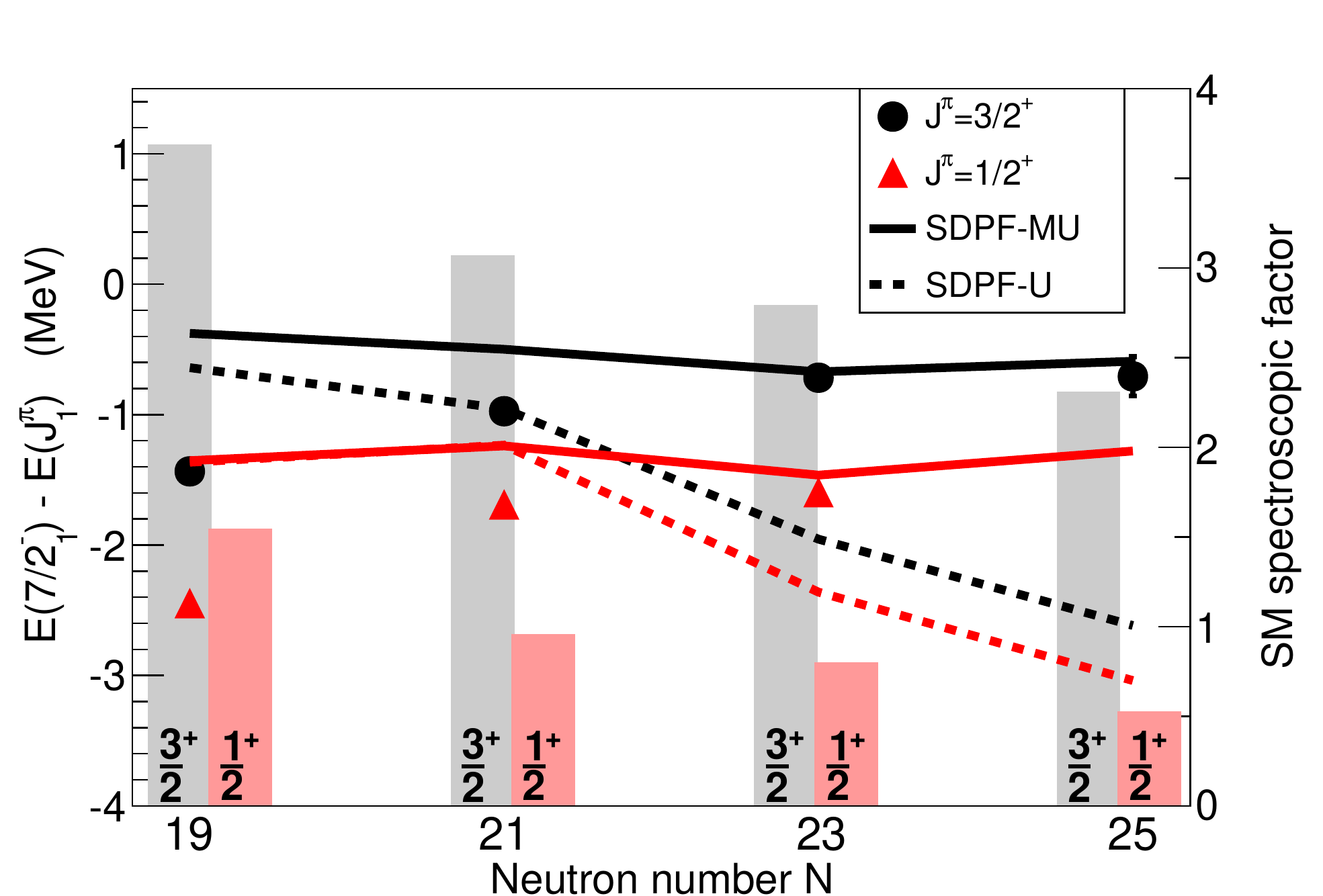}
\caption{(color online) Energies of $3/2^+$ and $1/2^+$ intruder states in
the silicon isotopes as a function of mass number. The points indicate
experimental data, while the lines connect theoretical calculations. The
vertical bars show the shell-model spectroscopic factors of the states.}
\label{fig:monopoles}
\end{figure}

Finally, the newly-measured energies of the $3/2^+$ and $1/2^+$ hole states
provide guidance for shell-model effective interactions that include excitations
across the $N=20$ shell gap. Figure \ref{fig:monopoles} shows the experimental
energies of the $3/2^+_1$ ($1/2^+_1$) states relative to the $7/2^-_1$ states,
indicative of the $f_{7/2}$ to $d_{3/2}$ ($s_{1/2}$) shell gap. For reference,
we also show the shell-model spectroscopic factors for populating these states
by one-neutron removal. The experimental data indicate that both gaps shrink as
neutrons are added from $N=19$ to 25, while SDPF-MU predicts a flat trend and
SDPF-U predicts an increase of these gaps.

\begin{figure}
\includegraphics[width=1.0\columnwidth]{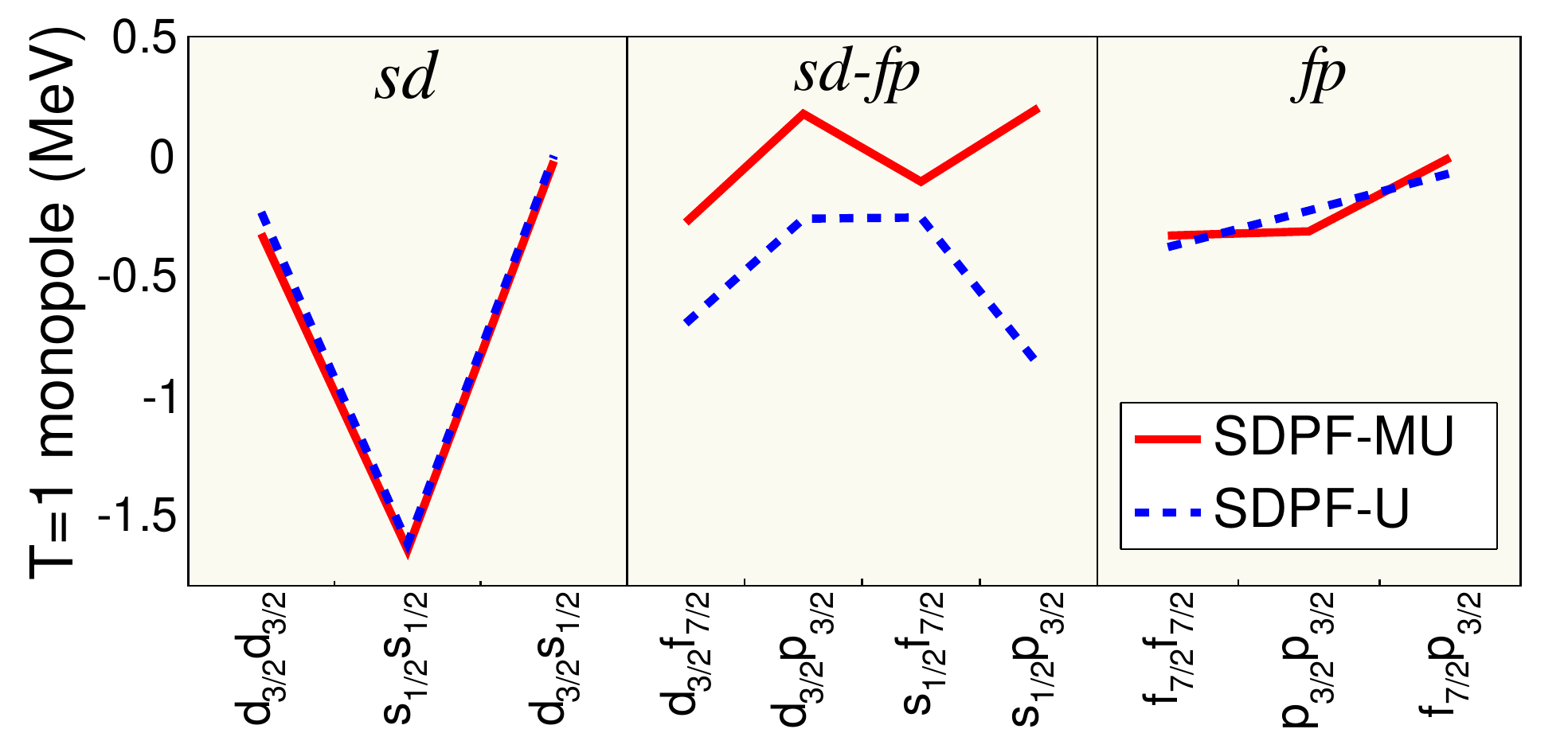}
\caption{(color online) Comparison of selected $T=1$ monopoles terms of the SDPF-U and SDPF-MU
effective interactions, evaluated at $A=42$.}
\label{fig:T1Monopoles}
\end{figure}

These qualitiatively different predictions can largely be attributed to a difference in the
cross-shell neutron-neutron ($T=1$) interaction.
Figure \ref{fig:T1Monopoles} shows
selected monopole (\ie angle-averaged) terms of the SDPF-U and SDPF-MU interactions.
While both interactions have similar $sd$ and $fp$ monopoles and are successful in reproducing the
spectroscopy of the region within the $0\hbar\omega$ model space, the more-attractive SDPF-U
cross-shell monopoles over-bind the neutron $sd$ orbits as neutrons are added to the $fp$ shell,
leading to the observed trend.
This discrepancy highlights a key difference between the two interactions.
In SDPF-U, due to insufficient experimental data, the cross-shell part of the interaction
was left as essentially the two-body $G$ matrix.
On the other hand, the cross-shell component of SDPF-MU was generated from the
schematic potential $V_{MU}$ \cite{Otsuka2010} which allowed---by incorporating
 information from data closer to stability---the approximate inclusion of the
repulsive contribution of three body forces to the effective $T=1$ two body interaction.
This same repulsive $T=1$ effect has been shown to be robust consequence of the Fujita-Miyazawa process
which is crucial in reproducing the oxygen dripline \cite{Otsuka2010a}.
We note that a more recent version of SDPF-U \cite{Caurier2014}, developed to allow neutron
excitations across the $N=20$ gap, in fact produces significant improvement over the original
SDPF-U energies \cite{PovesPC}.

In conclusion, we have exploited one-neutron knockout reactions to probe
the evolution of the $f_{7/2}$ and $p_{3/2}$ spectroscopic strength in
neutron-rich silicon isotopes. State-of-the-art shell-model interactions
describe the trends of the data but underestimate the role of the $p_{3/2}$
orbital. We confirm that the tensor force is necessary to describe the
evolution of the $f_{7/2}$ strength, and show that it is already
important at $N=26$. The observed excess of $p_{3/2}$ strength relative
to shell-model predictions indicates that the $N=28$ shell gap may be
reduced even more than present calculations suggest. Neutron cross-shell
excitations across the $N=20$ shell gap were identified and quantified
for the first time from the observation of positive-parity final states.
The shell-model interactions considered (SDPF-U and SDPF-MU) over-predict
the measured $d_{3/2}$ and $s_{1/2}$ neutron removal yields, pointing to
the deficiency of the applied model space truncations.
We have also identified the energies of neutron-hole states which depend
strongly on previously unconstrained neutron-neutron monopole interactions.
A comparison of shell-model predictions indicates the importance of three-body
forces in the evolution of structure in this region.

\begin{acknowledgments}
We thank the staff of the Coupled Cyclotron Facility for the delivery of 
high-quality beams. We also thank A. Poves for helpful
discussions. This material is based upon work supported by the 
Department of Energy National Nuclear Security Administration under Award 
Number DE-NA0000979. This work was also supported by the National Science 
Foundation under Grant No. PHY-1404442 and by the United Kingdom Science 
and Technology Facilities Council (STFC) under Grants ST/J000051/1 and 
ST/L005743/1. GRETINA was funded by the US DOE - Office  of Science. 
Operation of the array at NSCL is supported by NSF  under Cooperative 
Agreement PHY-1102511(NSCL) and  DOE  under grant DE-AC02-05CH11231(LBNL).
\end{acknowledgments}

\bibliographystyle{apsrev}
\bibliography{library,Priv_comm}{}

\end{document}